\documentclass[useAMS,usenatbib,usegraphicx,amsmath]{mn2e}
\title[Dynamics of Jupiter Trojans during the
2:1 mean motion resonance crossing of Jupiter and Saturn]
{Dynamics of Jupiter Trojans during the 
2:1 mean motion resonance crossing of Jupiter and Saturn}

\author{F. Marzari $^1$, H. Scholl $^2$}
\author[F. Marzari, P. Tricarico, H. Scholl]
{F. Marzari$^{1}$\thanks{E-mail:
marzari@pd.infn.it}
and H. Scholl$^{2}$\\
$^{1}$Department of Physics, University of Padova, Via Marzolo 8,
35131 Padova, Italy\\
$^{2}$Observatoire de la C\^ote d'Azur, BP 4229, 06304 Nice
CEDEX 4, France}
\begin{document}

\date{Accepted .... Received .....; in original form ...}

\pagerange{\pageref{firstpage}--\pageref{lastpage}} \pubyear{...}

\maketitle
\label{firstpage}

\begin{abstract}
\noindent
In the early phase of the Solar System evolution,
while the outer planets migrated due to their interaction with a planetesimal
disk, Jupiter may have crossed the 2:1 mean motion resonance with Saturn.
It is well known \citep{morna} that this dynamical event has profound consequences
on the evolution of an alleged initial Trojan population of Jupiter.
In this paper, we analyse
in details the dynamics of Jupiter Trojans during the resonance crossing.
We find that orbital instability is not confined to the central 2:1 resonance
region 
but occurs in a more extended region where a secular and secondary resonances 
perturb the Trojan orbits while the planets approach, cross and leave the 
2:1 resonance. 
In addition, Jupiter and Saturn are locked after the resonance crossing
in an apsidal corotation which has an additional destabilizing effect
on Trojans. The synergy of the secular resonance, secondary resonances
and apsidal corotation is needed to fully remove an initial Trojan population. 
New Trojans can be temporarily
captured from the planetesimal disk while Jupiter crosses this 
extended instability region. After the disappearence of major secondary
resonances, the secular resonance and the break of the apsidal corotation,
the temporarily captured Trojans are locked and can remain stable
over long timescales.
\end{abstract}

\begin{keywords}
celestial mechanics -- asteroids.
\end{keywords}

\section{Introduction}

According to a widely accepted scenario, the outer planets of the
solar system are embedded in a gas-free disk of planetesimals
in the last stage of planetary formation. Gravitational interactions
between planets and planetesimals dominate the dynamical evolution
of the disk. Planetesimals are scattered by the planets in a chaotic manner.
Orbital angular momentum and energy are exchanged resulting in 
planetary migration eventually with significant orbital changes.
This phenomenon, described extensively for the first time by Fernandez and 
Ip (1984),
was invoked then to suggest Pluto's capture in a 3/2 resonance by
Neptune during its outward migration \citep{malo}.

At what heliocentric distances did the planets form and how
far did they migrate? Two major models are proposed: 
in the first model investigated by Fernandez and Ip (1984) and applied by 
Malhotra (1993),
initial planetary orbits are widely spaced between 5.2 and about 25 AU.
A recently proposed model, the NICE model \citep{tsi}, assumes an initially
closely spaced distribution between 5.3 and 17 AU. In the latter model,
Uranus and Neptune exchange their orbits during migration.

The two migration models differ mainly in the assumption of the
initial semimajor axes of the planets. The driving mechanism for
planetary migration is the same: planets scatter planetesimals
in-and outwards. Inwards scattering moves a planet outwards while
it moves inwards when a planet scatters a planetesimal outwards. In a closed
system in equilibrium without loss of planetesimals where planets
scatter planetesimals in- and outwards, no significant migration would occur.  
In an open system with loss of planetesimals, on the other hand, significant planetary 
migration is possible. Jupiter plays a crucial role since it ejects easily
planetesimals received from the other planets out of the solar system.
As a consequence, Jupiter migrates towards the Sun while the other three planets
migrate outwards. Migration is halted when the outermost planet reaches
the edge of the planetesimal disk and when most of the planetesimals
scattered between the planets are removed. 

Jupiter and Saturn cross in the NICE model the 2:1 mean motion resonance (MMR)
soon after planetary migration has started.
As a consequence, the eccentricities of both planetary orbits increase.
Saturn approaches the orbit of the third planet which is excited and which,
therefore, has close approaches with the fourth planet. The third and fourth planet
may exchange orbits which moves the third planet rapidly towards 20 AU deep inside
the planetesimal disk surrounding in the beginning the four planets. 
Dynamical friction with planetesimals damps rapidly enough
planetary eccentricities to avoid close encounters between the third
and fourth planet which would result 
eventually in a destabilization of the outer 
planetary system. The planetary orbits separate due to migration and
their eccentricities are damped to present values due to dynamical friction. 

In a scenario where Jupiter and Saturn cross the 2:1 MMR, 
Jupiter Trojans are destabilized. 
The destabilization was first attributed to the particular perturbations
of the 2:1 MMR solely \citep{mibero}. Later, Morbidelli et al. (2005) attributed the
destabilization to the 3/1 secondary resonance between harmonics of
the libration frequency of
Trojan orbits and a critical argument of the 2:1 MMR.
This secondary resonance is very effective to remove Trojans
in case of a very low migration speed. Within
a frozen model without migration, all Trojans are removed \citep{morna}
on a timescale of about 1 Myr. 
One cannot
exclude, however, that a considerable fraction of Trojans survives since
each secondary resonance is quite narrow and Trojans may pass through. 
In this paper, we will show that
due to the presence of a major secular resonance on both sides of the 2:1 MMR,
original Trojans are removed due to the synergy between
secondary and secular resonances independently of the planet migration rate. 
In addition, the lock of Jupiter's and Saturn's
orbits into apsidal corotation after the 2:1 MMR crossing 
significantly contributes to
the destabilization until the locking is broken.

While primordial Trojans are destabilized before, during and after 
the crossing of the
2:1 MMR, nearby planetesimals can be
temporarily trapped on Trojan orbits via the reverse
chaotic path. As soon as Jupiter
leaves the extended instability region, the latest captured Trojans
remain locked on tadpole orbits for long timescales comparable to the age
of the planetary system. 
Morbidelli et al. (2005) have shown
that the orbital distribution of the observed Trojans corresponds to the
orbital distribution of the captured Trojans.

Temporary trapping in coorbital motion  appears to occur still at present.
\cite{eve} described temporary captures in horseshoe orbits and 
\cite{kar} identified about 20 transitional objects in a sample of 
about 1200 Trojans. Candidates are Centaurs that can be trapped
as Trojans for short periods of $10^4-10^5$ yrs \citep{hoev}. 
This shows that the stable region 
for Jupiter Trojans is  
surrounded by a chaotic layer \citep{marscho} where 
a population of temporary Trojans resides.  
At present,
the stable and unstable regions are well separated
and an object residing in the transient population 
cannot become a permanent Jupiter Trojan without the help 
of a non-conservative process. There are some slow diffusion gates 
from the stable to the unstable region like those identified 
by \citep{roga} related to commensurabilities between the secular 
frequency of the Trojan perihelion longitude and the 
frequency of the Great Inequality (2:5 almost resonance
between the present Jupiter and Saturn). However, it is 
very unlikely that a transient Trojan can follow in the 
reverse sense these paths 
to became a permanent Trojan. 

We describe in section 2 the major perturbations acting near the 2:1
MMR on Jupiter
Trojans in the early phase of the NICE migration model.
Section 3 is devoted to the synergy of these major perturbations
leading to a total loss of a possible initial Trojan population independent
of migration rates.
In section 4 we show that perturbations in the central Jupiter-Saturn 
2:1 MMR region,            
where at least one of the resonant arguments librates,
do not lead to global instability even in a frozen model as
suggested by Michtchenko et al. (2001). A Frequency Map Analysis 
reveals extended stable regions.

\section{Sources of instability for Jupiter Trojans 
before, during, and after the 2:1 MMR crossing}

In this Section we describe the sweeping of resonances through the
Trojan region before and after
the 2:1 MMR crossing
of Jupiter and Saturn. 

\subsection{The numerical models}

The goal of our numerical modellig is to explore the stability 
of Trojan orbits during the migration of Jupiter and 
Saturn through the 2:1 MMR. The migration rates of the two 
planets have to be computed within a model that includes all 
the outer planets and a disk of massive planetesimals as described in 
\citep{tsi}. For this reason we have first reproduced 
the dynamical evolution of the outer planets using the same model 
of \citep{tsi} and adopting the same SYMBA5 
numerical algorithm \citep{dunc_levi}. 
It is a symplectic integrator that models the gravitational 
interactions among planets, the gravitational
forces exerted by the planets on planetesimals and vice versa. The gravitational
interactions among planetesimals is omitted in order to gain computing time.
SYMBA5 is particularly
designed to handle close encounters among planetesimals and planets, 
the main mechanism driving the migration of the outer planets. 
Using the starting conditions for the planets described in 
\citep{tsi} for relatively slow migration, 
we performed a numerical simulation that
matches closely that shown in \citep{tsi}. Hereinafter, we 
refer to this simulation as PPS (Planets and Planetesimals Simulation). 
The four outer planets are started on closely packed,
almost circular and
coplanar orbits. The semimajor axes $a$ of Jupiter and Saturn are
5.45 and 8.50, respectively, so that they will cross the 2:1 MMR during
their migration. Orbital eccentricities $e$ 
and inclinations $i$ are equal to 0.001 at start.
Following \citep{tsi}, we use 4500 massive planetesimals to 
produce the migration of the
four planets.  

\begin{figure}
 \includegraphics[width=8.5cm]{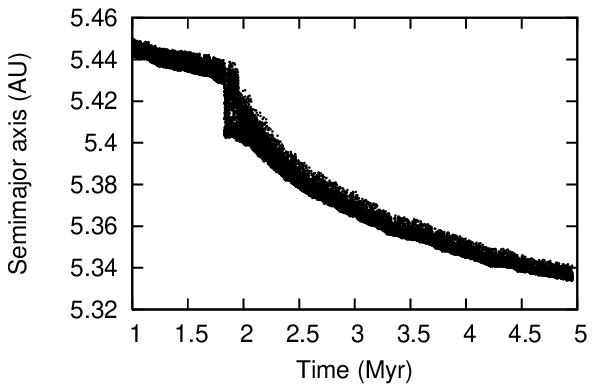}
 \includegraphics[width=8.5cm]{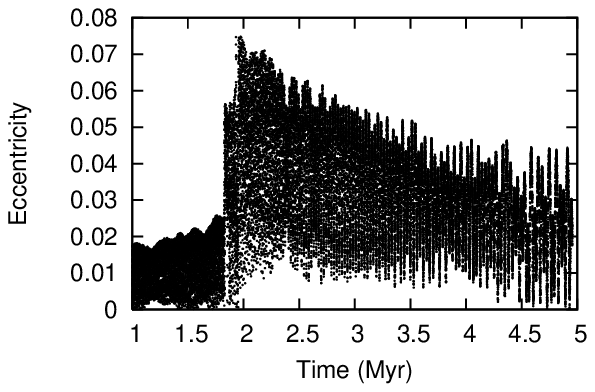}
\caption[]{Orbital evolution of Jupiter semimajor
axis and eccentricity during  the migration of the
planets
driven by planetesimal scattering (PPS simulation). 
The outcome is
very similar to that shown in \cite{tsi}. 
We focus on the resonance crossing that occurs after about 1.8 Myr
from the beginning of the simulation. 
}
\label{juppa}
\end{figure}

In Fig.~\ref{juppa} we show the semimajor axis and eccentricity 
of Jupiter as obtained in our PPS simulation. 
Before the 2:1 MMR crossing, Jupiter's eccentricity is equal on average
to 0.01 in spite of its small starting value. This is due to the 
forced component of Saturn which grows while approaching the 
resonance location. The 2:1 MMR crossing is  
marked by a sudden jump in eccentricity related 
to the separatrix crossing and by large
oscillations in semimajor axis. 
After the crossing, 
the eccentricity is slowly damped down while the planet continues to migrate
towards its present location. 

From the time series of the orbital elements of both Jupiter and 
Saturn, computed within the PPS simulation, we can derive the 
planet migration 
rate $da/dt$ of the semimajor axis and the eccentricity damping 
rate $de/dt$ and produce 
a {\it synthetic} model. In this model the effect of
the planetesimal scattering is simulated by adding analytically 
the $da/dt$ and $de/dt$ terms 
to the equations of motion of the planets. Such an approach
was exploited to model 
the effect of circumstellar
disks on exoplanets, for instance by \citet{lpe} and \citet{kle}.
The authors used analytic expressions to estimate the changes 
in $a$ and $e$ due to the interactions with the disk
when advancing the planets from
time $t_i$ to $t_{i+1}$.  
We follow the formalism outlined in the appendix of the paper by \citet{lpe}
and, to model the migration of planets,
we introduce a $da/dt$ and $de/dt$ in the SYMBA5 integrator 
to produce the migration and neglect all the massive planetesimals. 
We concentrate on the orbital evolution of Jupiter and Saturn 
since thay are responsible for the stability
or instability of Jupiter Trojans. 
Uranus and Neptune are needed in the PPS model 
in order to transport the planetesimals responsible for the migration
of the outer planetary system. However, by a series of numerical tests, we
have verified that their influence on the Trojan orbits of 
Jupiter is negligible compared to that of Jupiter itself
and Saturn. 

In the synthetic model 
we must account for the fact that 
the migration of 
Jupiter and Saturn caused by planetesimal encounters
is linear only over a limited amount of time and not
over the whole migration period. 
The number of planetesimals in planet 
crossing orbits is in fact declining causing a slow decrease of 
$da/dt$ and $de/dt$.  We, therefore, tune the synthetic integrator 
to the PPS run by using values of $da/dt$ and $de/dt$ 
that are derived from PPS at different times during the 
evolution of Jupiter and Saturn.  
In this way, the synthetic model accurately reproduces
the evolution of the planets during the 2:1 MMR and even after. Moreover, 
it retains all the dynamical features needed to analyse the stability
of Jupiter Trojans. The initial values of $da/dt$ and $de/dt$ for Jupiter
are $-7.39 \times 10^{-9}$ AU/year and $-3.76 \times 10^{-10}$ 1/year,
respectively. After 10 Myr these values have decreased to 
$-4.05 \times 10^{-9}$ AU/year and $-2.05 \times 10^{-10}$ 1/year.
For Saturn, the $da/dt$ ranges from $2.23 \times 10^{-8}$ 
to $1.24 \times 10^{-8}$  AU/year in 10 Myr while the $de/dt$ goes from 
$-2.80 \times 10^{-9}$  to  $-1.65 \times 10^{-9}$ 1/year.

The main advantage of using the synthetic integrator is its speed. 
We can compute the orbital evolution of Jupiter and Saturn and of 
massless Trojans on a timescale at least 100 times shorter than 
that required by a full model that includes the massive 
planetesimals (PPS--type model). CPU time is a critical issue since 
we have to explore the stability of Trojans in the 
phase space for different intervals of time during migration 
and in different dynamical configurations. In addition, with the 
synthetic model there is the possibility of easily changing the 
values of $da/dt$ and $de/dt$ which are strongly model dependent. 
We tested also a synthetic model based on the RADAU 
integrator and the results were in agreement with the 
SYMBA5  synthetic model. 

To identify possible resonances between the motion of
the planets and that of Trojans we have to evaluate
the major orbital frequencies of these bodies. However, the 
dynamical system evolves because of planetary migration 
and the frequencies change with time. To compute the 
value of these frequencies at a given instant of time 
we use {\it frozen} models. We extract the osculating orbital elements 
of the planets and Trojans at the required time and start a 
numerical integration of the trajectories with the migration 
switched off (both $da/dt$ and $de/dt$ are set to 0). 
In this way, we compute a time series of  
orbital elements for the non--migrating planets and Trojans 
long enough to derive precise values of 
the frequencies.  

To compute initial orbital elements for Trojans 
at different times during the evolution of the planetary system
we select random initial conditions within a ring surrounding 
the orbit of Jupiter. The semimajor axis of any putative
Trojan is selected in between $0.9$  $a_J$ and 
$1.1$ $a_J$ where $a_J$ is the semimajor axis of 
Jupiter. The eccentricity can be as large as $0.5$ and the inclination
extends up to $50^{\circ}$. The other orbital angles are selected
at random between $0$ and $360^{\circ}$. Each set of initial conditions
is integrated for $10^4$ years and if the critical argument 
$\lambda - \lambda_J$ librates in this time period, 
a body with that set of initial 
conditions is included 
in the sample of virtual Jupiter Trojans. The choice of wide ranges in 
eccentricity and libration amplitude, somewhat wider than the 
present ones,  is dictated by the 
chaotic evolution of the orbital elements before, during and after the 
2:1 MMR crossing. This chaotic evolution can drive a given Trojan orbit
from a high eccentric orbit into an almost circular one, and it can 
strongly reduce its libration amplitude. We cannot neglect at this 
stage orbits that are unstable on the long term since they might 
be turned into stable ones during the dynamical evolution 
caused by the planetary migration. A body is considered to be 
ejected out of the swarm during its evolution when its critical 
argument no longer librates. 

\subsection{Secular resonance with Jupiter}

\begin{figure}
 \includegraphics[width=8.5cm]{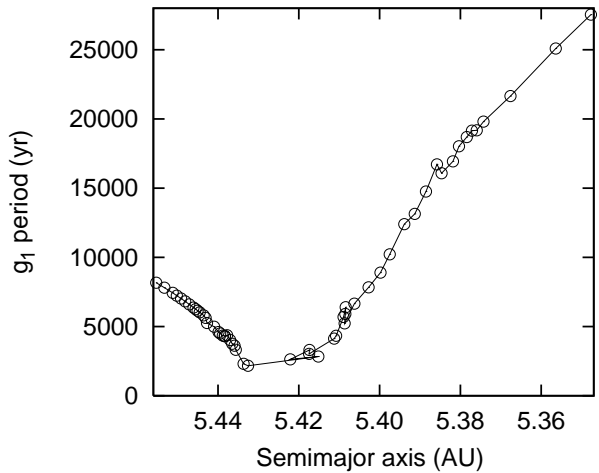}
\caption[]{Period corresponding to frequency $g_1$ as a function
of Jupiter's semimajor axis.
}
\label{g1_vs_a}
\end{figure}

The secular evolution of eccentricities and 
perihelion longitudes of the Jupiter--Saturn system,
as described by the Lagrange--Laplace averaged theory,
is characterized
by two major frequencies that we call $g_1$ and $g_2$ following 
Murray and Dermott (1999).  
These frequencies are not constant during 
planetary migration since their values depend on the 
semimajor axes of the two planets through the Laplace coefficients. 
The linear Lagrange--Laplace theory 
has an analytical solution that allows to compute 
both $g_1$ and $g_2$ as a function of planetary orbital elements. 
However, this solution fails in proximity of the 2:1 MMR
and we resort to 
a full numerical approach to compute the two frequencies
during planetary migration. 
We 'freeze' the dynamical system at different stages of 
migration (frozen model) and we estimate both $g_1$ and $g_2$ from 
the time series of the non--singular variables $h$ and $k$
of Jupiter over $1 \times 10^6$ yrs. As usual, we define these variables by
$h = e* cos(\varpi)$ and $k = e* sin(\varpi)$.
For the computation of precise values for the two frequencies 
we use the so-called Modified Fourier Transform (MFT) analysis 
\citep{la1,la2,la3},
which we had already applied to study the stability properties
of the 
present Jupiter Trojan population
\citep{marscho}. $g_1$ and $g_2$ are by far the frequencies 
with the largest amplitude computed from the MFT. 

\begin{figure}
 \includegraphics[width=8.5cm]{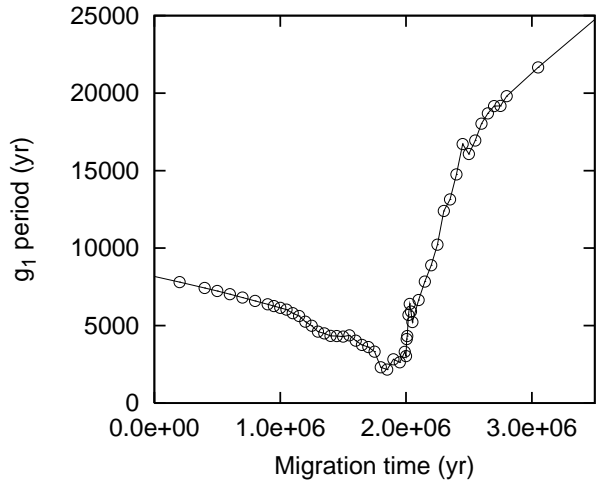}
\caption{Period corresponding to frequency $g_1$ as a function
of time.
}
\label{g1_vs_time}
\end{figure}

\begin{figure}
 \includegraphics[width=8cm]{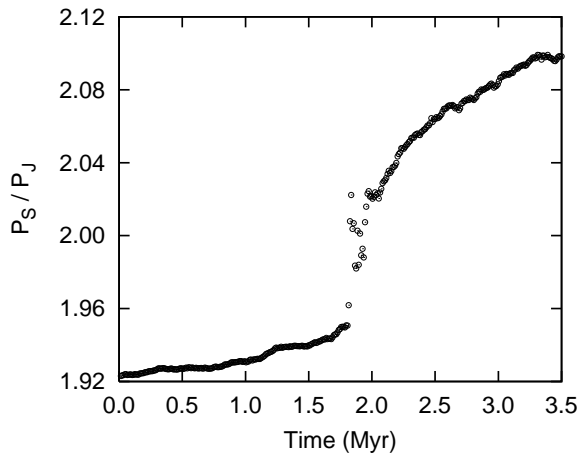}
\caption{Ratio between Saturn's ($P_S$) and Jupiter's ($P_J$) orbital
periods as a function of time.
}
\label{ratio_P}
\end{figure}

\begin{figure}
 \includegraphics[width=8cm]{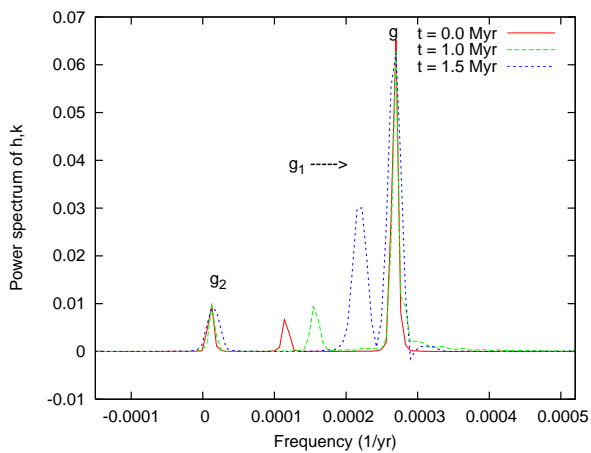}
\caption{Power spectrum of $h$ and $k$ for a Trojan while Jupiter and
Saturn approach the 2:1 MMR. The secular frequencies $g_1$ and $g_2$
of the Jupiter-Saturn system and the Trojan's
proper frequency $g$ are given at three different
times. Frequency $g_1$ moves towards $g$ destabilizing the Trojan's
orbit at 1.5 Myr.
}
\label{power_spec_trojan}
\end{figure}

One of the two frequencies sweeps through the Trojan 
region during the migration of the planets reaching 
values typical of  the proper frequency
$g$ of Jupiter Trojans. We call this frequency 
$g_1$ while the other frequency, $g_2$,
has a longer period and does not influence 
the Trojan motion. 
When $g_1$ is equal or very close to $g$ 
a secular resonance is established.
Fig.~\ref{g1_vs_a} and Fig.~\ref{g1_vs_time} show the behaviour
of $g_1$ 
as a function of Jupiter's semimajor
axis and of time during migration, respectively.
The period corresponding to $g_1$ decreases while Jupiter and Saturn approach 
the 2:1 MMR  and it rises back after the 2:1 MMR.
Fig.~\ref{ratio_P} shows for
comparison the ratio of the orbital periods between Saturn, $P_S$, and
Jupiter, $P_J$.
When Jupiter and Saturn approach, cross and leave the 2:1 MMR, 
$g_1$ sweeps through the Trojan phase space causing strong perturbations
that lead mostly to instability.
Libration amplitudes and/or eccentricites of Trojans are
increased resulting in close encounters with Jupiter.
Due to the functional dependence of $g$ on the proper 
elements of the Trojan orbits \citep{marscho}
the secular resonance appears first
at high inclinations, moves then down to low inclinations 
when the planets reach the 
2:1 MMR, and finally climbs back to high
inclinations after resonance crossing. This behaviour 
will be described in more detail in Section 3.1.
Fig.~\ref{power_spec_trojan} shows a power spectrum of the complex
signal $h + i k$ for a Trojan orbit. The frequency 
$g_1$ approaches $g$ when the planets migrate towards the 
2:1 MMR leading to resonant perturbations.
The frequency $g_2$
does not change much and remains far from $g$. 
The Trojan becomes unstable just after the third instant of time
as shown in Fig.~\ref{power_spec_trojan}
before the 2:1 MMR is crossed (t=1.5 Myr). 
When it falls inside the $g = g_1$ resonance its orbit is 
in fact destabilized on a short timescale by a fast change in 
eccentricity and libration amplitude. 

However, the delicate dynamical equilibrium of the 
Trojan motion is perturbed even when 
$g_1$ is only close to $g$, outside the 
secular resonance borders. 
The term proportional to $g - g_1$ in the disturbing function 
is dynamically important generating a chaotic evolution of 
Trojan orbits even if on a longer timescale compared
to those cases falling into the resonance. 
A similar effect was observed for Uranus 
Trojans whose diffusion 
speed in the phase space is strongly increased, leading to
chaotic motion, in proximity of the fundamental frequencies
$g_5$ and $g_7$ of the solar system \citep{mara}. 
When $g_1$ leaves the Trojan region after the 
2:1 MMR, it remains anyway close to $g$ 
for a long time persisting as a source of instability.
Moreover, after the 2:1 MMR, Jupiter and Saturn are locked in
an apsidal resonance that enhances the strength of the 
$g - g_1$ term by
coupling the perturbations of Jupiter to those of Saturn
(see next section). 
 
A change in the initial values of Jupiter and Saturn in the migration model 
would move the location of the 2:1 MMR and the 
corresponding values of 
the semimajor axes of both Jupiter and Saturn
at the crossing. 
However, this does not alter the 
effect of the secular resonance on the stability of Trojans.
The resonance sweeping 
occurs anyway since $g_1$ and $g_2$ depend on 
the semimajor axis of Jupiter $a_J$, according to  
the Lagrange--Laplace averaged theory, in the same way as
the frequency $g$ depends on $a_J$
following Erdi's theory of Trojan motion \citep{erdi}.

\subsection{Secondary resonances with harmonics of the '2:1 Great Inequality'}

There are two independent
critical resonance arguments for the 2:1 MMR of Jupiter and Saturn:
$\theta_1 = \lambda_J - 2 \lambda_S + \varpi_J$ and 
$\theta_2 = \lambda_J - 2 \lambda_S + \varpi_S$, 
where $\lambda$ and 
$\varpi$ denote respectively mean longitude and
longitude of perihelion. Either one of the two critical arguments
librates while the other circulates or both critical arguments 
librate simultaneously.
In the latter case, the difference between the two critical
arguments 
$\theta_2 - \theta_1 = \Delta \varpi$ also librates. This means that
Jupiter and Saturn are in apsidal corotation.

\begin{figure}
 \includegraphics[width=8.5cm]{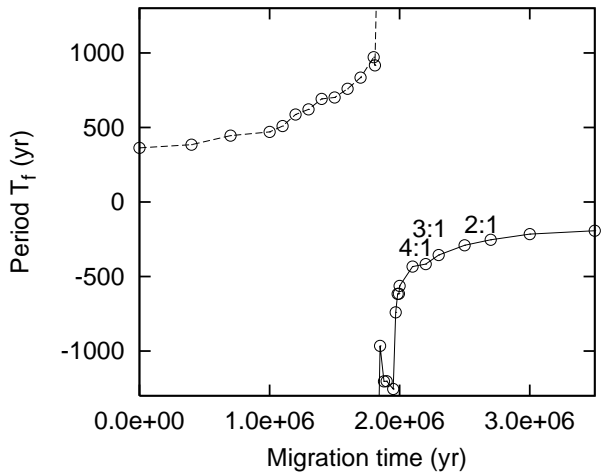}
\caption{Evolution of the circulation period $T_f$ of $\theta_1$ as
a function of time.
The resonance crossing is marked by
a discontinuity in the period of $\theta_1$.}
\label{period_theta}
\end{figure}

While Jupiter migrates towards the Sun and Saturn in opposite direction,
both $\theta_1$ and
$\theta_2$ circulate prograde before the
2:1 MMR and retrograde after. 
The frequency of $\theta_1$ and $\theta_2$ may become
commensurable with the libration frequency of the 
critical argument of Jupiter Trojans.
This is the case of a secondary resonance
which was  investigated by Kortenkamp et al.(2004) for Neptune Trojans. The authors found
that a Neptune Trojan in a secondary resonance can enhance significantly its 
libration amplitude possibly leading in some cases to instability. 
The importance of secondary resonances for Jupiter Trojans in the frame of 
the NICE model was recognized by Morbidelli et al. (2005). 
Secondary resonances can
be encountered before and after the crossing of the 2:1 MMR.
In a frozen model without
migration, the 3:1 secondary resonance after the 2:1 MMR 
removes all Trojans on a timescale of 1 Myr while the
2:1 secondary resonance removes $70\%$ of them. In a migration model, these
removal rates can be significantly less if the secondary resonances
are crossed rapidly.

\begin{figure}
 \includegraphics[width=8.5cm]{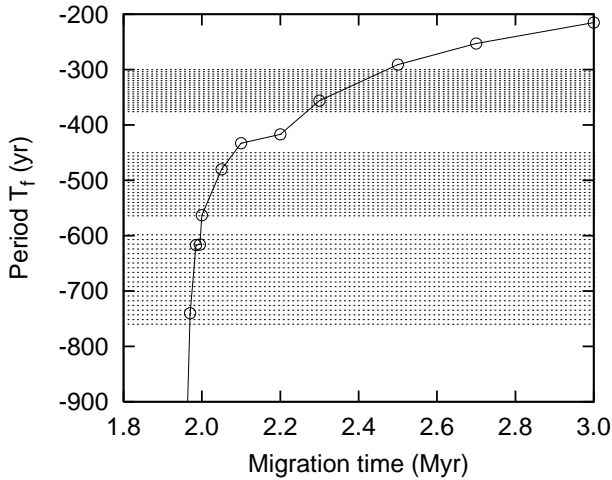}
\caption{Crossing of the 4:1 (lower shaded stripe), 
3:1 (middle stripe) and 3:1 (upper
stripe) secondary resonances.}
\label{size_second_reso}
\end{figure}

In Fig.~\ref{period_theta} we show the period $T_f$
of the frequency $f$ of circulation of $\theta_1$ as a
function of the semimajor axis of Jupiter during migration.
Different secondary resonances are crossed.
Crossing, however,  is fast, in particular after 
the 2:1 MMR. In proximity of the 4:1 secondary resonance,
for example, the period of $\theta_1$ changes by approximately 20\% in 
only $3 \times 10^4$ yrs. In Fig.~\ref{size_second_reso} we
illustrate with a shaded stripe the frequency interval of
$f$ (translated into periods) for which there is
a  4:1  (lower shaded stripe), 3:1 (middle shaded stripe) and 
a 3:1 commensurability (upper shaded stripe) with the 
libration frequency of a Trojan swarm. We consider Trojans up to 
$50^{\circ}$ in inclination and up to 0.35 in
eccentricity corresponding to libration periods roughly
ranging from 145 and 190 yrs. The sweeping appears to be fast, 
in particular for the 4:1 and 3:1 secondary resonances, taking also 
into account that any individual Trojan will be affected 
only by a fraction of the time spent by $f$ to cover 
the entire shaded region. It is worthy to note here that the 
migration speed is relatively low within the different NICE models
\citep{tsi,morna}. 
A faster migration would further reduce the relevance of secondary
resonances in the destabilization of Jupiter Trojans during the 
2:1 MMR.  

\begin{figure}
 \includegraphics[width=8cm]{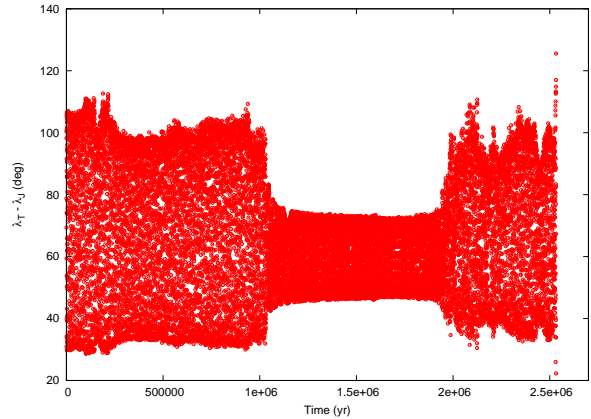}
\caption{Jumps in the libration amplitude of a Trojan while
crossing the secondary resonances 2:1 near 1 Myr and 3:1
near 2 Myr before the 2:1 MMR.
The Trojan enters at the end near 2.5 Myr the secular resonance $g_1$ and is
destabilized.
}
\label{secondary_res_crossing_before_MMR}
\end{figure}

As for the secular resonance, the crossing of the secondary resonances 
occurs before and after the 2:1 MMR. However, there is a substantial 
difference between the two dynamical configurations. Before the 2:1 MMR the 
secondary resonance sweeping causes sharp jumps in libration amplitude and 
eccentricity that in most cases do not fully destabilize the 
Trojan orbit. As shown in 
Fig.~\ref{secondary_res_crossing_before_MMR} the crossing of the 
2:1 secondary resonance at $t \sim 1 Myr$ reduces the libration 
amplitude increasing the stability of the orbit. When the 3:1 secondary
resonance is encountered later, the initial libration amplitude
is restored. The Trojan orbit becomes finally unstable when it 
crosses the secular resonance with $g_1$. Of course, for librators
with large amplitudes,
the perturbations of the secondary resonances may lead to a
destabilization of the Trojan orbit. 

Totally different is the dynamical behaviour after the 2:1 MMR.
The secondary resonances are much more effective in destabilizing 
Trojan orbits independent of their libration amplitude. 
The reason for the different efficiency of secondary 
resonances  before and after 
the 2:1 MMR is due to two independent causes:
\begin{itemize}

\item Immediately after the 2:1 MMR crossing, 
the eccentricity of Jupiter is on average 
higher. This reinforces presumably secondary resonances and the 
secular resonance. 
We tested this hypothesis before the 2:1 MMR by numerically 
integrating the same Trojan orbits in a model with the 
eccentricity of Jupiter set to an average value 2.5 times 
higher compared to that of the 
reference model which approximately corresponds to the 
average increase observed in simulations. 
Trojans are started
in between the 2:1 and 3:1 secondary resonance. Trojans surviving at least 
$2 \times 10^4$ years decreased  by 33 \% with respect to the 
Jovian low eccentricity  case. On average, the 
lifetime was reduced by 22 \% in the higher eccentric case.

\item After the 2:1 MMR crossing the planets are always locked in 
apsidal corotation, according to our simulations. This additional 
dynamical effect contributes to destabilize Trojan orbits. 
To estimate the effects of apsidal corotation we have 
used the same model described in the previous item (that with
higher eccentricity) and forced
apsidal corotation of the planets before the 2:1 MMR crossing
by a convenient choice of the orbital angles of the planets. 
A comparison between apsidal and non-apsidal corotation
after resonance crossing is not possible since the system finds always
rapidly the apsidal corotation state.
In the apsidal corotation model the number of surviving Trojans 
drops by about
23 \%, compared to that without apsidal corotation, 
and the Trojan lifetime is shortened by 42 \%.
\end{itemize}

\subsection{Effect of apsidal corotation between Jupiter and Saturn
on the dynamics of Trojans}

\begin{figure}
 \includegraphics[width=9cm]{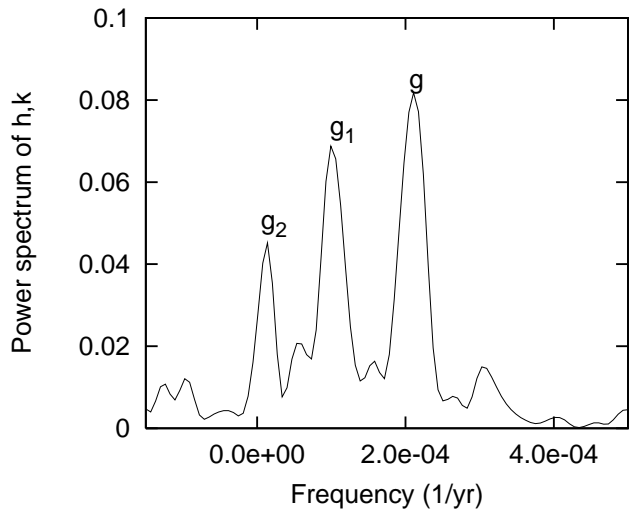}
\caption{Power spectrum of a Trojan trajectory after the
2:1 MMR crossing. The apsidal corotation increases the strength
of the secular term $g_1$ that leads to chaotic evolution.
}
\label{power_spectrum_after}
\end{figure}

\begin{figure}
\hskip -1truecm
 \includegraphics[width=8cm,angle=-90]{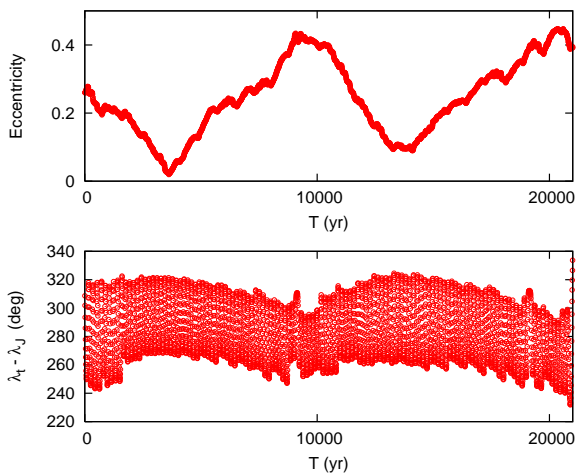}
\caption{Orbital evolution of a  Trojan after the
2:1 MMR crossing perturbed by the 4:1 secondary
resonance. The libration center is shifted from $300^\circ$
to a lower value due to the apsidal corotation between Jupiter
and Saturn.
}
\label{4_1_sec_reso_after_cross}
\end{figure}

After the 2:1 MMR crossing, Jupiter and Saturn are locked in 
apsidal corotation in all our simulations. In most cases, apses are
antialigned with $\Delta \varpi = \varpi_J - \varpi_S$ librating
about $180^\circ$.
This apsidal corotation is broken much later.
The presence of apsidal corotation, as stated in the
previous section, has significant consequences 
for the instability of Trojans:
\begin{itemize}
\item It enhances the effects of the $g - g_1$
secular term  since the frequencies of the precessional
rates for the perihelia longitudes of Jupiter and Saturn
are commensurable.
In Fig.~\ref{power_spectrum_after} 
we show the power spectrum of a Trojan started
in between the 2:1 MMR and the 3:1 secondary resonance. The two
peaks corresponding to the $g_1$ and $g_2$ frequencies of the 
Jupiter and Saturn system are clearly visible and $g_1$ is close to the 
proper frequency $g$. The peaks are much higher as compared to
the power spectrum in Fig.~\ref{power_spec_trojan} which is obtained before
the 2:1 MMR where $\Delta \varpi$ circulates.  

\item The secondary resonances after the 2:1 MMR become very effective  in 
destabilizing Trojan orbits due to the increased eccentricity of Jupiter 
as pointed out above.  
The coupling between the apsidal corotation and secondary resonances 
causes a 
fast growth of the eccentricity and a corresponding
shift in the libration center of the Trojan tadpole motion away from 
the Lagrangian points L4 and L5 \citep{chris}.
In  Fig.~\ref{4_1_sec_reso_after_cross} 
we show an example for this shift and eccentricity
increase
in the 4:1 secondary resonance after the 2:1 MMR crossing. 
The Trojan is destabilized at the end by a close encounter with Jupiter.
Note that this behaviour is never 
observed {\it before} the 2:1 MMR crossing where there is no 
apsidal corotation and the eccentricity of Jupiter is lower. 
\end{itemize}

\section{Synergy between secondary resonances, the secular resonance $g_1$
and apsidal corotation}

In order to investigate the combined effects of the three identified
major perturbations, we start Trojan populations 
at different migration stages before and after the 2:1 MMR crossing.
Simulations begin in between the major secondary resonances 4:1,
3:1, 2:1, 1:1, and right before and after
the onset of apsidal corotation.

The starting values are produced by generating randomly
the initial orbital elements and checking for the 
critical libration angle.
Maximal starting inclinations and eccentricities are taken somewhat 
larger than 
in the presently observed Trojan population of Jupiter because, 
as already anticipated in Section 2.1, the chaotic evolution
in the proximity of the resonance
crossings may reduce their values and lead to a stable tadpole orbit.

\begin{figure}
 \includegraphics[width=8cm]{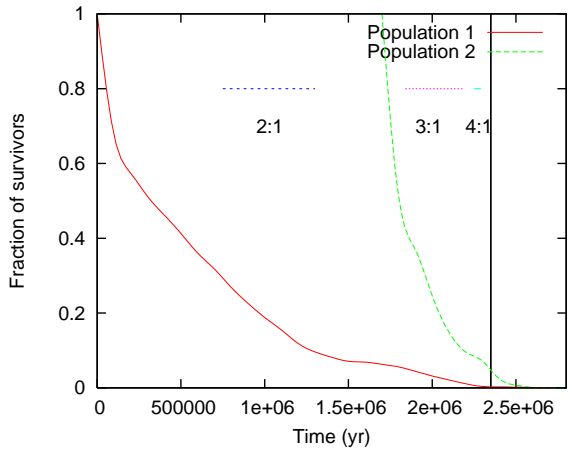}
\caption{Evolution of two fictitious Trojan populations while
Jupiter and Saturn approach the 2:1 MMR. Population 1
is started 2.3 Myr before the resonance crossing. It is
eroded mainly by the secular resonance $g_1$. Secondary resonances
affect solely librators with large amplitudes.
Population 2 is started $6 \times 10^5$ yrs before the 2:1 MMR.
Somewhat more bodies survive since the sweeping $g_1$ resonance
affects less high inclined Trojans.
Behind the 2:1 MMR, these
surviving bodies are removed by effects discussed below.
The continuous line marks the 2:1 MMR. The dotted horizontal 
lines show the location of the secondary resonances over all 
the range of libration amplitude of the bodies in the 
two populations. The timespan covered by the secondary
resonances is shrinking because the migration is faster
in proximity of the 2:1 MMR.
}
\label{erosion_two_populations}
\end{figure}

\subsection{Fate of Trojans before the 2:1 MMR crossing}

When Jupiter and Saturn approach the 2:1 MMR, Trojans cross 
secondary resonances and, in particular, the $g_1$ secular resonance.
Secondary resonances before the 2:1 MMR are a very weak 
instability source and 
destabilize solely tadpole orbits with large libration amplitude. 
The secular resonance
$g_1$, on the other hand, may remove Trojans with any libration amplitude
when it sweeps through the region. Even when a body is not 
exactly within the $g_1$ resonance but closeby it feels the perturbations 
of the $g - g_1$ term and it may be destabilized, even if on a longer
timescale. 
Fig.~\ref{erosion_two_populations} illustrates the erosion of two initial 
Trojan populations starting at different times, 2.3 and respectively 0.6 Myr 
before the 2:1 MMR. The two populations are generated with the same 
random process described in Section 2.1 and, as a consequence, they 
are dynamically similar. 
It appears that the secular resonance and the secondary 
resonances are more effective close to the 2:1 MMR where they 
destabilize more than 90\% of the Trojan population. The location 
of the secondary resonances are shown during the evolution of 
the populations as horizontal lines in 
Fig.~\ref{erosion_two_populations}. Before the 2:1 MMR resonance crossing, 
the sweeping of these resonances is slower than after the
the 2:1 MMR when the migration of the planets is much faster. 
However, as discussed 
above, before the 2:1 MMR 
secondary resonances are significantly 
weaker because of the reduced eccentricity of 
Jupiter. We also recall that each individual Trojan has 
its own libration period and it is affected by the 
secondary resonances only during a fraction of the time covered
by the resonance sweeping. 

\begin{figure}
 \includegraphics[width=9cm]{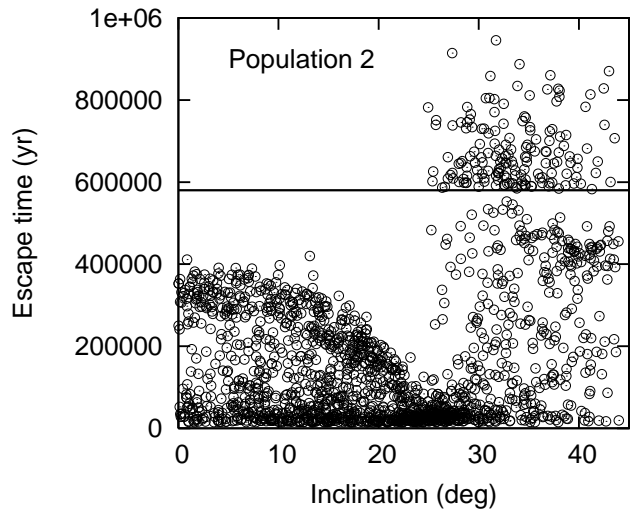}
\caption{Escape time for population 2 Trojans of
Fig.~\ref{erosion_two_populations}
vs. initial orbital inclination. The secular
resonance sweeps through the Trojan region starting from an
inclination of about $25^{\circ}$ down to low inclined
orbits. The continuous line marks the 2:1 MMR.
}
\label{escape_times_pop2}
\end{figure}

\begin{figure}
  \begin{center}
    \begin{tabular}{cc}
\includegraphics[width=2.7cm,angle=-90]{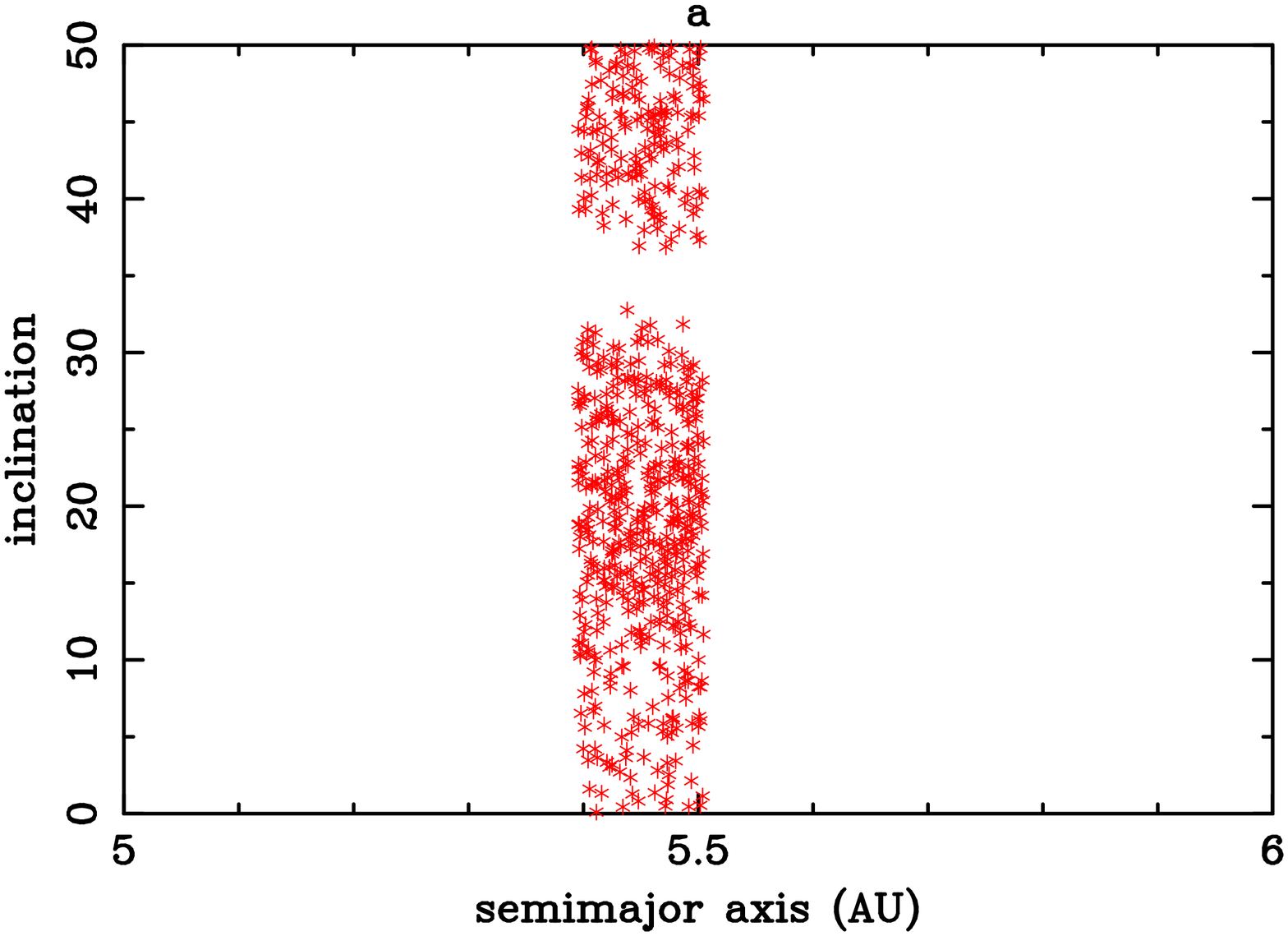} &
\includegraphics[width=2.7cm,angle=-90]{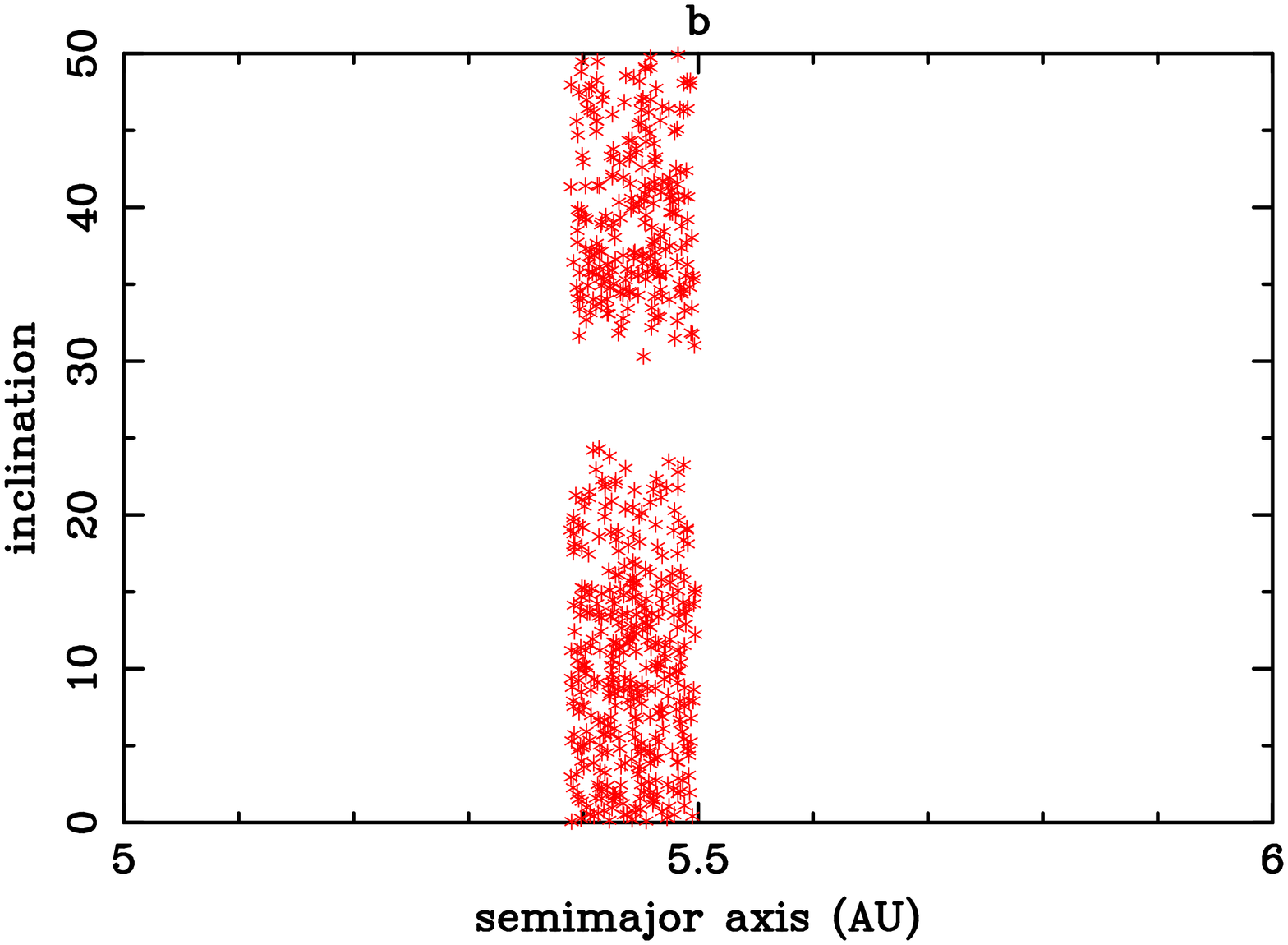} \\
\includegraphics[width=2.7cm,angle=-90]{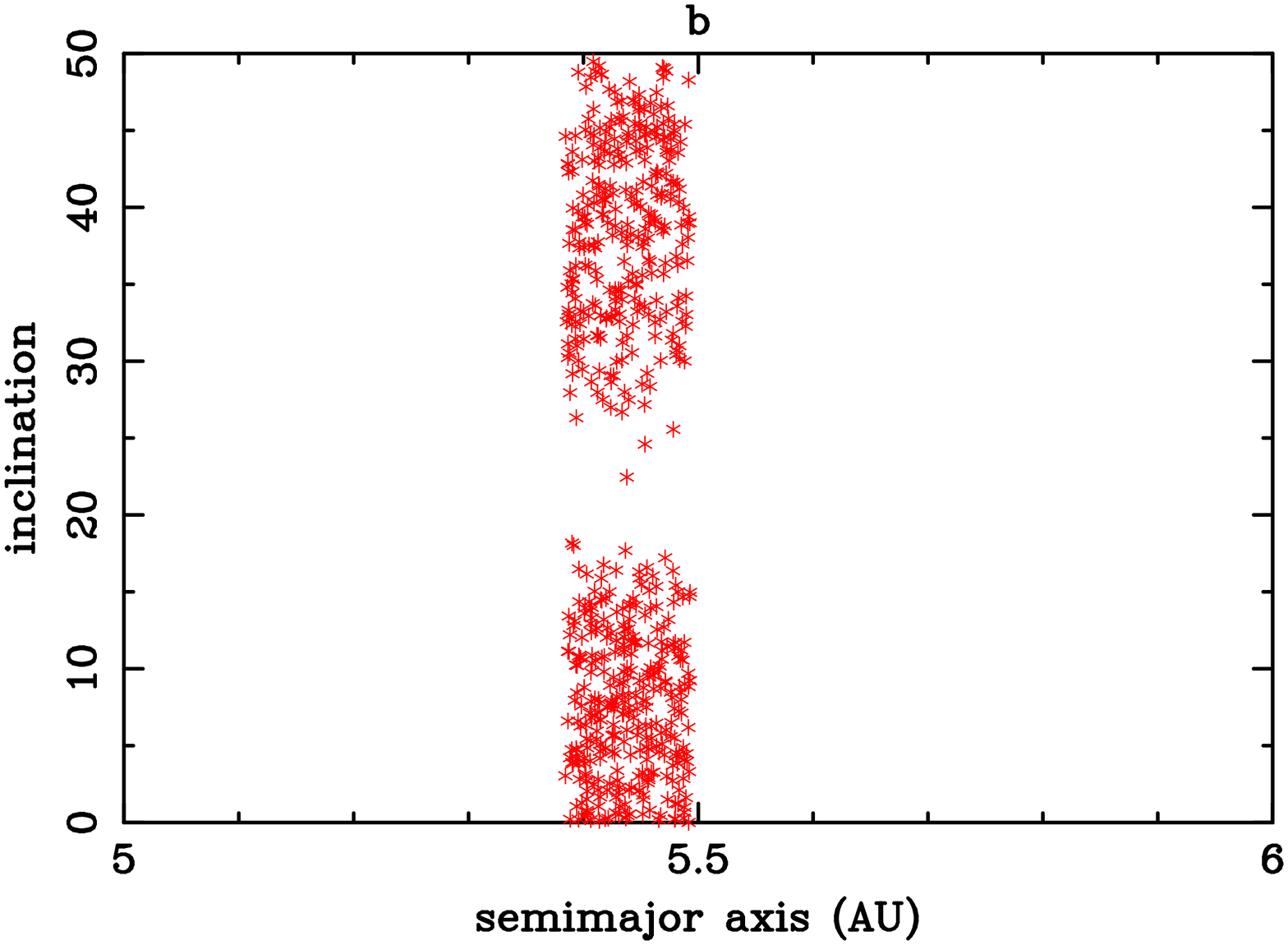}  &
\includegraphics[width=2.7cm,angle=-90]{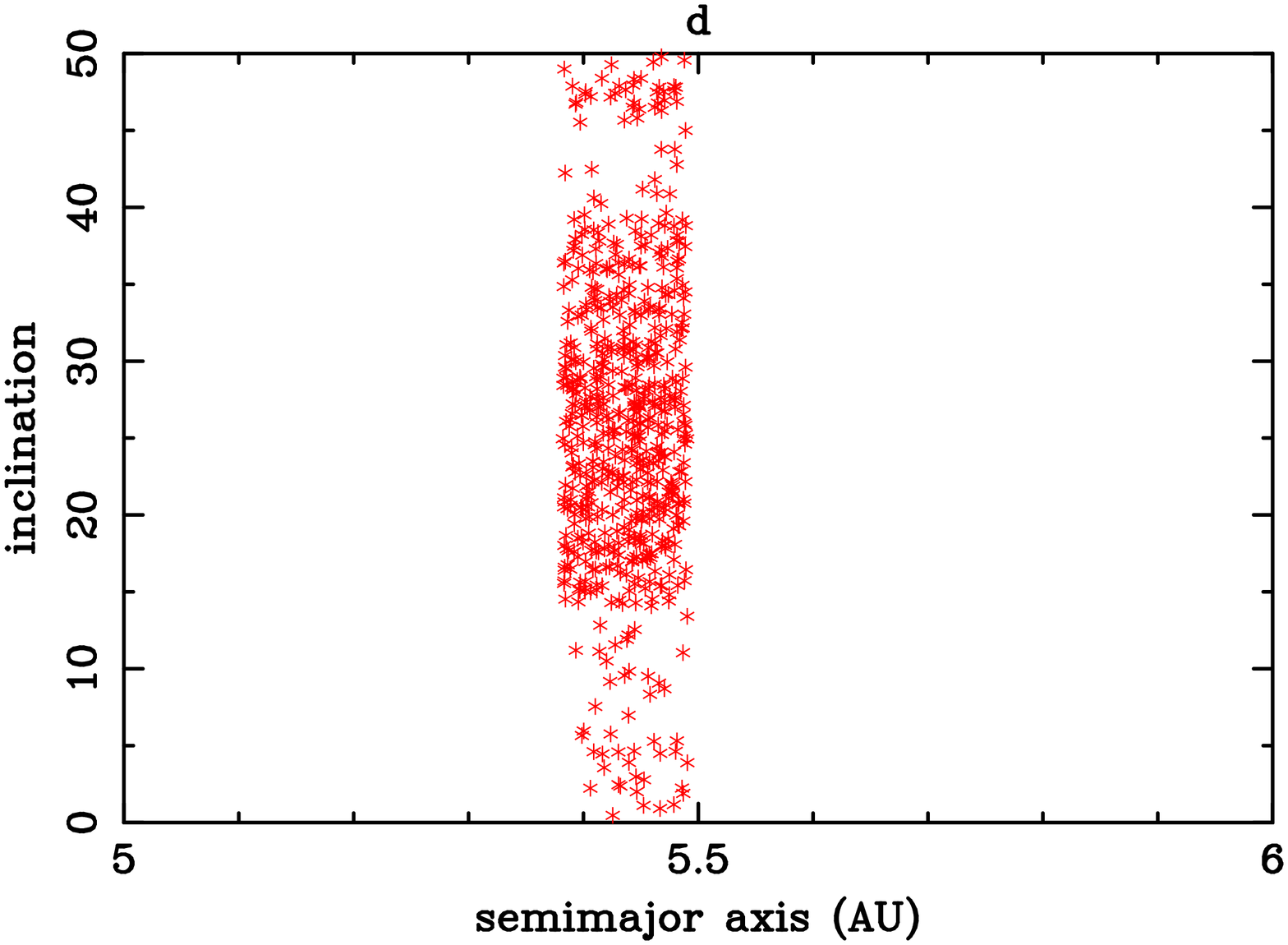} \\
    \end{tabular}
    \caption{Sweeping of the secular resonance when the
planets approach the 2:1 MMR. The orbits of the same
initial Trojan swarm
are integrated within 4 different frozen models
with the planets progressively closer to the
2:1 MMR.
}
    \label{escape_times_pop2_2}
  \end{center}
\end{figure}

The secular resonance sweeping through the 
Trojan region is
illustrated in Fig.~\ref{escape_times_pop2} and 
Fig.~\ref{escape_times_pop2_2}. 
Fig.~\ref{escape_times_pop2} shows the escape time
as a function of initial inclination for population 2.
The secular resonance at the beginning of the simulation 
destabilizes a large number of bodies with inclinations around $25^{\circ}$.
They all leave the Trojan region in less than $1 \times 10^5$ yr.
The critical value of $25^{\circ}$ for inclination is determined by the
choice of the initial orbits of Jupiter and Saturn before the 2:1 MMR
which in turn determines the value of $g_1$. A smaller initial 
distance between the planets, like in case of population 1, 
would have destabilized most Trojans at a higher inclination. 
As the planets move towards the 2:1 MMR during their migration,
the frequency $g_1$ increases and perturbs Trojan orbits at
a progressively lower inclination. According to Marzari et al.(2003), 
the proper 
frequency $g$ is higher for low inclined Trojans. 
In our sample of Trojans, there are naturally unstable orbits 
since we do not exclude those trajectories with large libration 
amplitude and high eccentricity. Some may be injected deeper in the stable region
after crossing secondary resonances.
Most of them, however, escape on 
a short timescale and populate the figure at the lower
edge of the y-axis. 

The high inclined Trojans that survive the 2:1 MMR crossing are destabilized 
when the frequency $g_1$ decreases again (its period grows) as shown in 
Figs.~\ref{g1_vs_a} and Figs.~\ref{g1_vs_time}. If some high inclination
Trojans survive somehow the first sweep of the secular resonance, 
either by chance or because the planets start their migration close
to the resonance location as in Fig.~\ref{escape_times_pop2}, they 
probably will
be destabilized by the second resonance 
sweeping when the planets move away from the resonance.

In Fig.~\ref{escape_times_pop2_2} we illustrate the distribution in 
inclination and semimajor axis of the same Trojan swarm integrated
in frozen models 
with the planets progressively approaching the 2:1 MMR. The empty stripe 
corresponds to the secular resonance destabilizing the orbits on a timescale 
of $1 \times 10^5$ yr in a frozen model. 
The sweeping proceeds towards lower inclinations 
while the planets approach the 2:1 MMR, whereas it rises back after
the crossing in a symmetric way. We like to emphasize that bodies in
the secular resonance are destabilized on a short timescale. 
Trojans whose frequency is close to $g_1$ but are not within the 
resonance borders are perturbed by the term $g -g_1$ and 
have a slower chaotic behaviour.
This explains why also high 
inclined Trojans in Fig.~\ref{escape_times_pop2} are slowly eroded away. 
It also accounts for the fact that both population 1 and 2 are fully 
destabilized notwithstanding that population 1 is started much farther 
away from the 2:1 MMR crossing. 

\subsection{Evolution of a Trojan population after the 2:1 MMR
crossing: from the 
4:1 to the 1:1 secondary resonance.}

\begin{figure}
 \includegraphics[width=8cm]{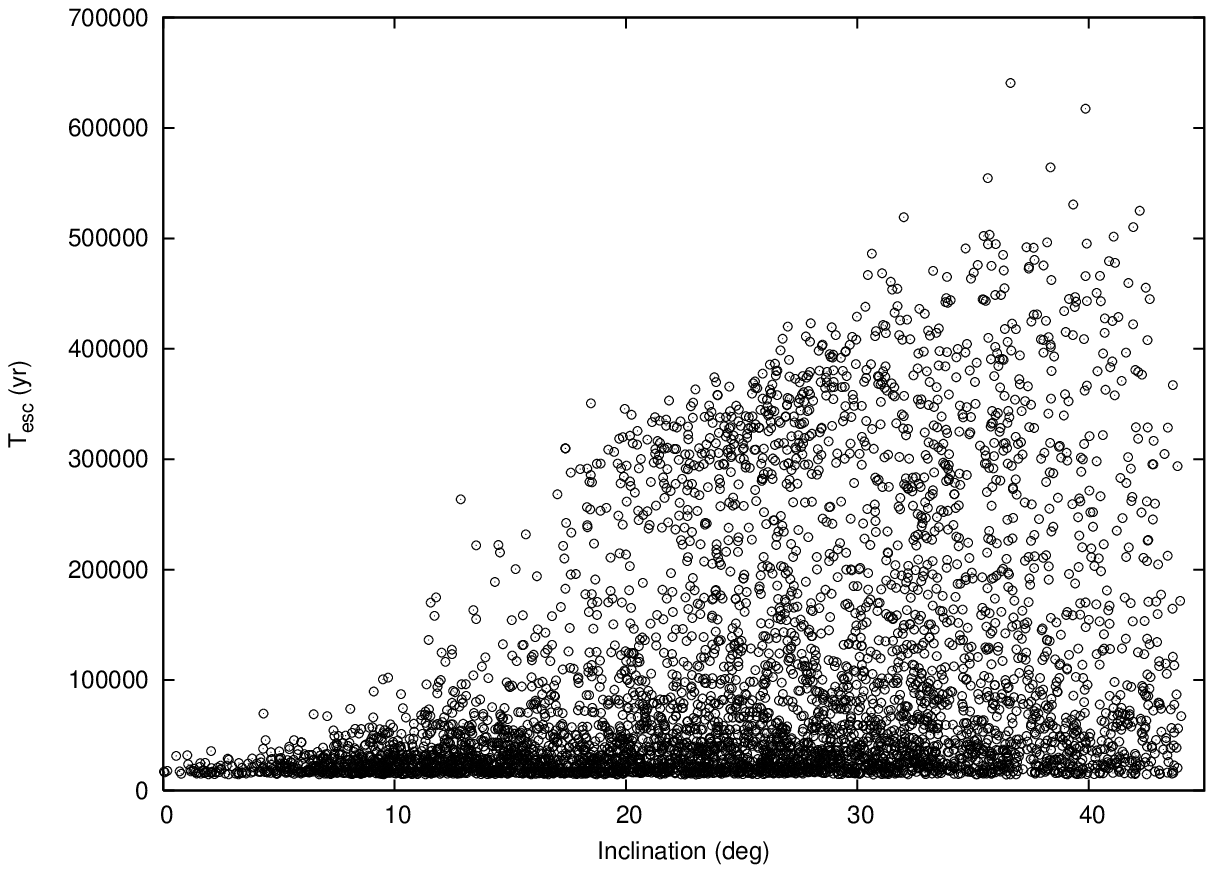}
 \includegraphics[width=8cm]{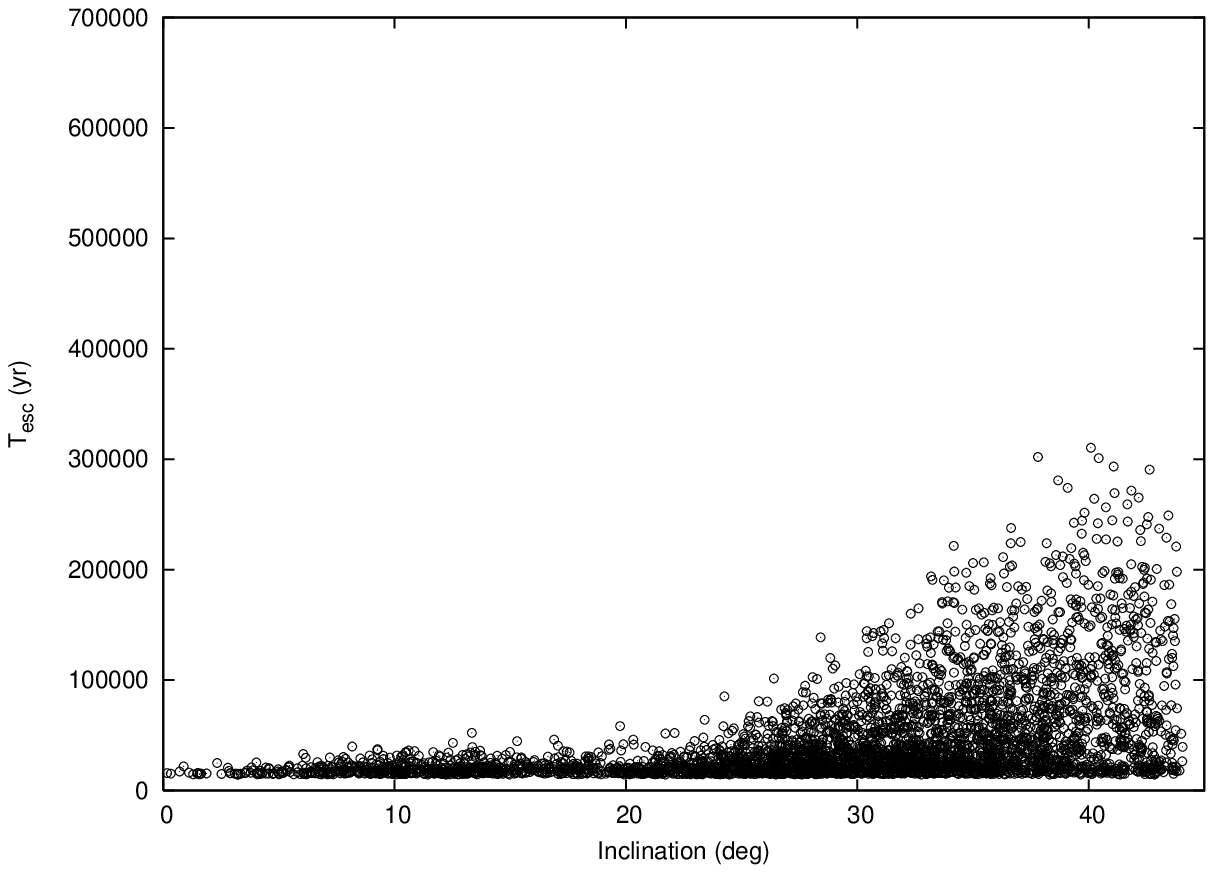}
\caption{Escape time  vs.
initial inclination for two
Trojan populations started after the 2:1 MMR crossing.
The first population (top) has libration frequencies
encompassed between the 4:1 and 3:1 secondary resonance.
The second population (bottom) is started
in between the 3:1 and the 2:1 secondary resonance.
}
\label{escape_times_incl}
\end{figure}

The closer Jupiter and Saturn start their migration to the 2:1 MMR,
the more initial Trojans may survive due to the dependence of
the destabilizing secular resonance on orbital inclination.
After the 2:1 MMR, the surviving Trojans would encounter 
a second time the secular resonance and the 
secondary resonances. Both the secular and the secondary 
resonances are reinforced
by apsidal corotation and by the larger eccentricity 
of Jupiter and 
the destabilization rate is significantly higher.
In Fig.~\ref{escape_times_incl} we show the evolution of two populations of 
Trojans after the 2:1 MMR. The population in the upper diagram
starts in between the 4:1 and 3:1 secondary resonance.
The Trojans of 
the second population in the lower diagram
have libration frequencies which place them
in between the 3:1 and 2:1 secondary
resonances. To understand the features of the two diagrams,
we have to keep in mind that 
the libration frequency depends on several orbital elements. A Trojan
population with about the same semimajor axis 
is spread over a large range of their other orbital parameters. As a consequence, 
the Trojans 
cross the same secondary resonance at different times. Moreover, 
the proper frequency $g$ depends mostly 
on the inclination and the effects of the secular resonance appear,
as already noted above in the discussion of Fig.~\ref{escape_times_pop2},
at different inclinations during the sweeping.
According to Fig.~\ref{escape_times_incl},  orbits with low inclination 
are rapidly destabilized while the escape time
grows significantly for inclinations higher than $15^{\circ}$. 
At high inclinations the secular resonance arrives
at a later time and it takes longer to destabilize Trojans.
Some of the orbits of the second population 
survive the 3:1 crossing but are ejected before reaching the 2:1
secondary resonance. Only a few Trojans
get beyond.   
At low inclinations where the 
instability is very fast, we observe a rapid pumping up of eccentricity 
and
a corresponding shift in the libration center. The power spectrum shows
also the vicinity
of the secular resonance. We conclude that, after the 
secondary resonance crossing, no initial Trojan
can survive due to 
the synergy between secondary and secular resonances. 
Their effects are
enhanced by the apsidal corotation of the two planets.  
The chaotic trapping of new Trojans appears to be difficult 
at low inclinations where 
the instability is very fast on timescales of the order of a few $10^4$
yrs, while it might be more efficient at higher 
inclinations where the slow instability might allow the formation of 
a steady state transient population of unstable Trojans. 

\begin{figure}
 \includegraphics[width=8cm]{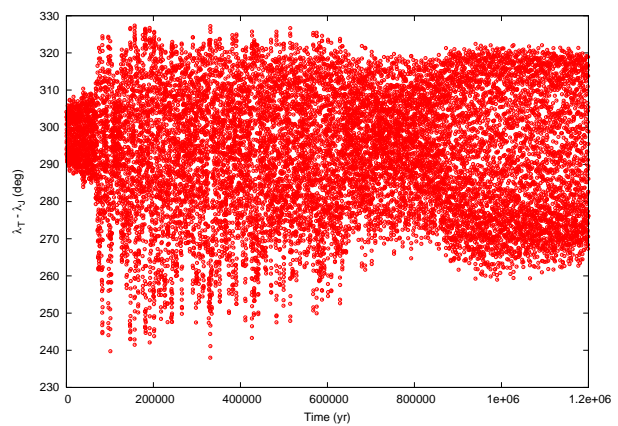}
\caption{Evolution of the critical argument of
a Trojan orbit crossing the 1:1 secondary resonance.
The crossing is marked by the first sharp jump in libration
amplitude. Subsequently, far from the secondary resonance, 
the orbit stays chaotic since the 
dynamics is still influenced by the
the secular frequency $g_1$ powered up by the 
apsidal resonance.
}
\label{1_1_secondary_res}
\end{figure}

At the 1:1 secondary resonance, significantly farther away from the 
2:1 MMR, a sharp jump in libration amplitude 
$D$ and eccentricity 
occurs for Trojan orbits. An example is given in Fig.~\ref{1_1_secondary_res} 
where $D$ changes 
during the 1:1 crossing. This resonance is weaker as compared to the 
previously encountered secondary resonances and it does not 
fully destabilize 
tadpole orbits but it induces chaotic variations of $D$. After the 
1:1 crossing, when the libration frequency is away from the 
frequency of either $\theta_1$ or $\theta_2$, 
the libration amplitude still shows an irregular behaviour.
By inspecting the power spectrum of the $h$ and $k$ variables 
of the Trojan orbit 
we find that the secular frequency $g_1$ is still relevant with a peak 
about half the size of the proper one. It is still a source of 
slow chaotic diffusion for the Trojan orbit.

\subsection{Far away from secondary resonances: still chaotic
changes of orbital elements}

\begin{figure}
 \includegraphics[width=8cm]{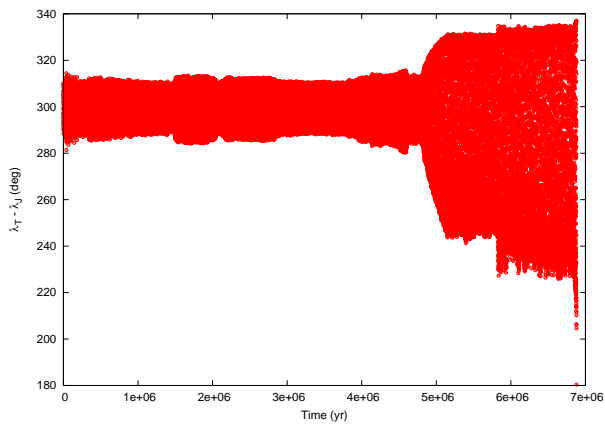}
\caption{The critical argument of a Trojan orbit
started far away from secondary resonances.
The orbit is still mildly chaotic because of the presence of the
secular frequency $g_1$. When Jupiter and Saturn cross 
the 4:9 mean motion resonance, the 
libration amplitude of the orbit increases
until ejection out of the Trojan region.
}
\label{far_away}
\end{figure}

As noted before, when the system gets beyond
the 1:1 secondary resonance
both the eccentricity and 
libration amplitude of tadpole orbits show a slow chaotic evolution
which is enhanced when the planets cross mutual higher order mean motion 
resonances. 
Fig.~\ref{far_away} shows the evolution of a
Trojan trajectory with initially small values of $D \sim 20^{\circ}$, 
$e_p \sim 0.03$ and $i_p \sim 19^{\circ}$. This orbit
would lie deeply in the stable region  
for the present configuration of the planets and it is 
far from any significant secondary resonance. 
The secular frequency $g_1$ appears to be still somehow
relevant for the stability of the Trojans causing moderate
libration amplitude variations. However, the orbit is 
finally destabilized during the crossing of a
4:9 mean motion resonance between Jupiter and Saturn.
The large libration amplitude increase 
beginning at $\sim 5 \times 10^6$ yr leads to a fast 
destabilization of the tadpole orbit. 
Significant changes of the proper elements are still possible 
at this stage and the
door for chaotic trapping is still open.

After the apsidal corotation is broken, 
and the weakened secular frequency $g_1$ has moved
farther away from $g$, 
Trojan orbits are finally stable on a long timescale with no 
detectable variations
of the libration amplitude. The door for chaotic capture is closed 
and the Trojan population approaches its present configurations with no
other significant remixing of proper elements. 

\section{When Jupiter and Saturn are in the 2:1 MMR} 

\begin{figure}
\hskip -1 truecm
\includegraphics[width=7.5cm,angle=-90]{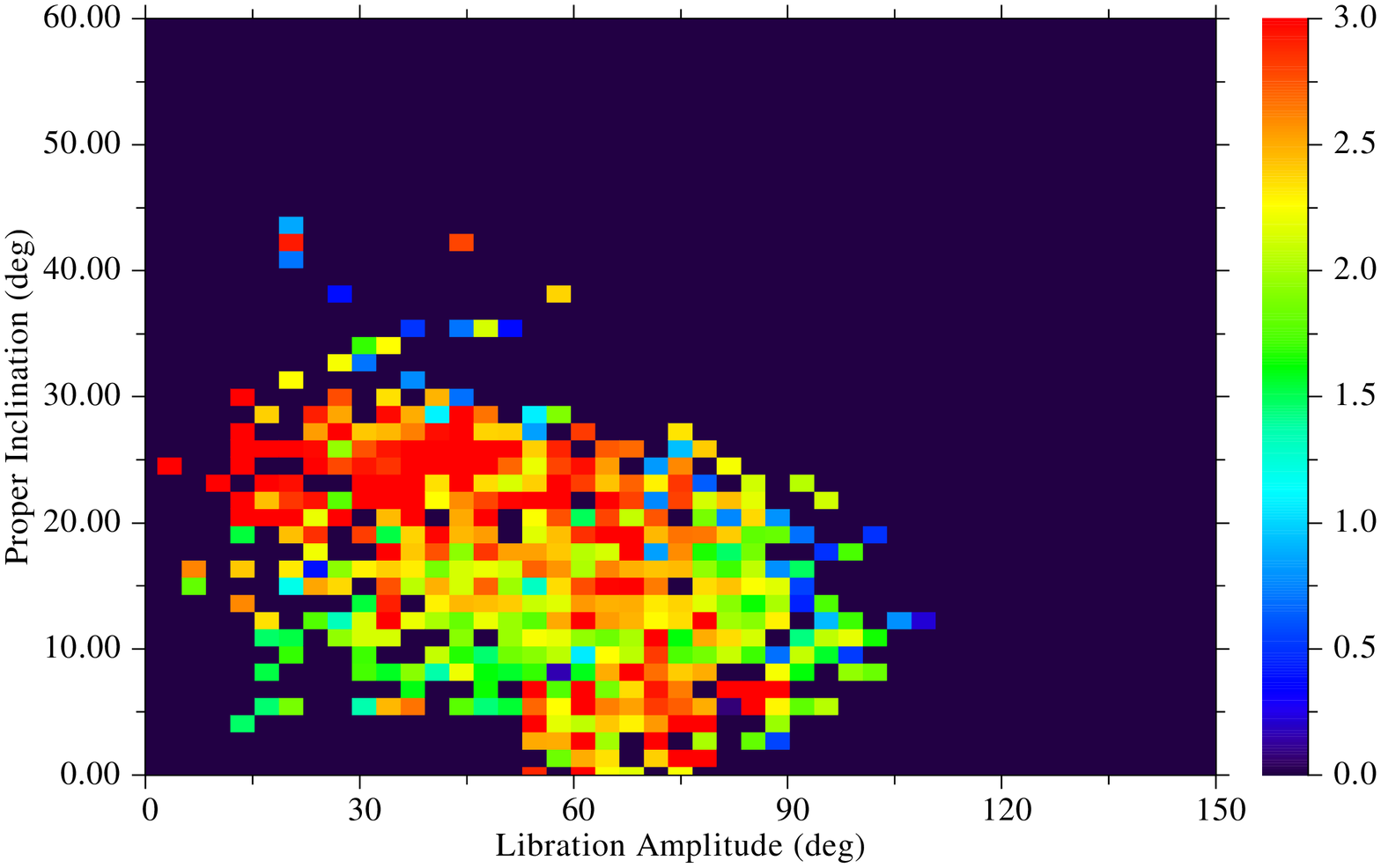}

\hskip -1 truecm
 \includegraphics[width=7.5cm,angle=-90]{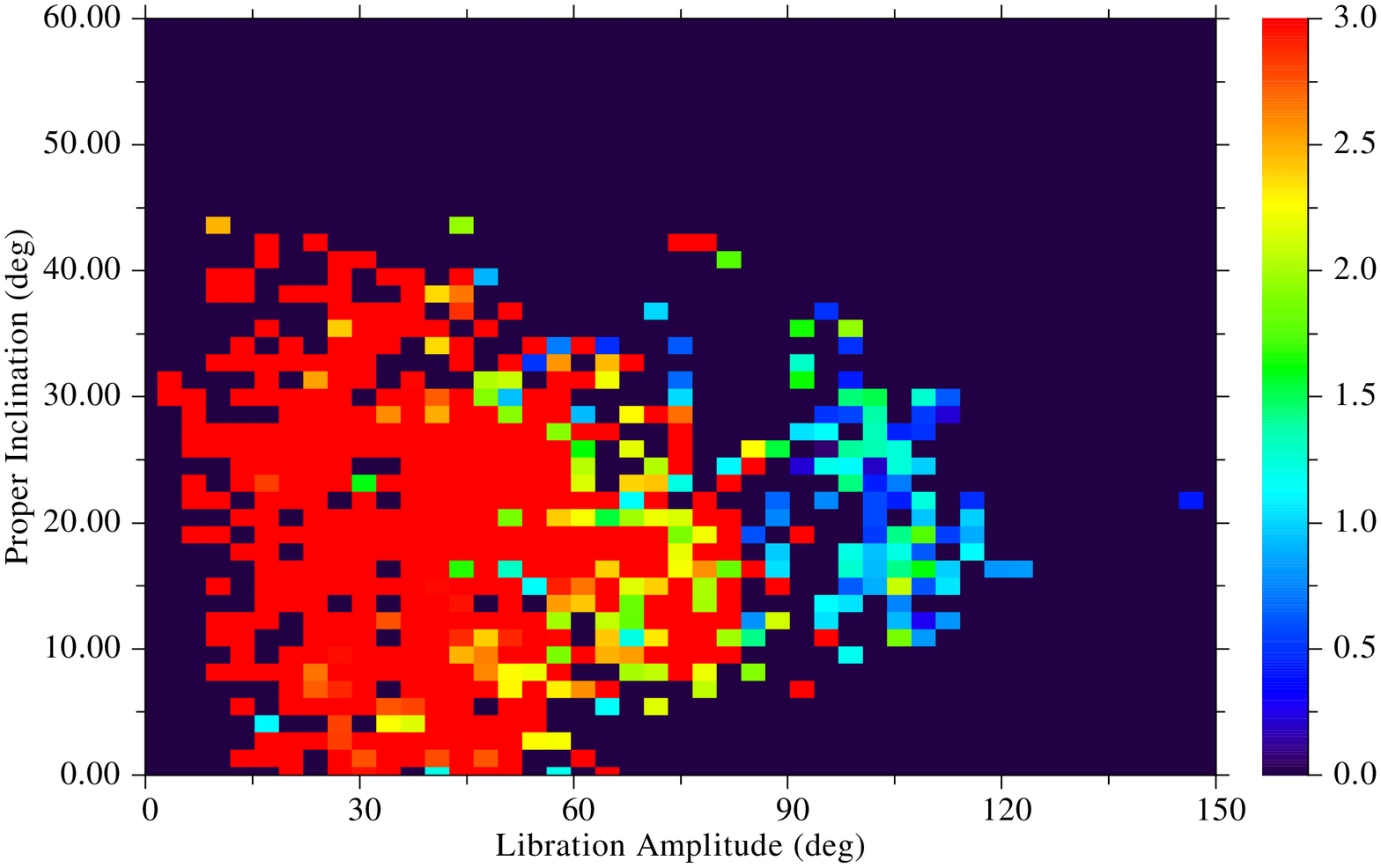}
\caption{Diffusion portraits of Trojan orbits for the 2:1 MMR.
In the upper diagram both
critical resonant arguments librate and
the planets are in corotation. In the lower diagram, where
a wider region has lower diffusion speeds, marked by
red squares, only one critical argument librates.}
\label{fig13a}
\end{figure}

The instability of Jupiter Trojans 
with the planets in the 2:1 MMR with Saturn was
investigated by \citet{mibero} in a frozen model. Jupiter and Saturn
are in apsidal corotation. Trojan starting values are confined to
inclinations of $5^{\circ}$, $\varpi - \varpi_J = 60^{\circ}$, 
$\lambda - \lambda_J = 60^{\circ}$, and eccentricity lower than
0.3. Using a RADAU integrator, the authors find instability over a very short
time scale of about $10^4$ yrs. This indicates that if the 
migration of Jupiter and Saturn was very slow, a temporary 
capture of the planets in the 2:1 MMR might have led to global 
instability of Trojans. However, when we performed 
numerical simulations of Trojan orbits  
in a frozen model
like Michtchenko et al.(2001) we did not
find short term instability. 
Using their semimajor axes for Jupiter and Saturn and confining
Trojans to their starting region, and using also a RADAU integrator,
we found a large number of stable Trojans over 
at least $10^5$ yrs. Instability
for this restricted starting region in phase space usually does not set on before
1 Myr. Similar results were obtained by \cite{nedo} and \cite{masb} in static
models were the planets were moving on fixed orbits. We will perform
here a more detailed analysis of the stability of Trojans 
when Jupiter and Saturn are locked in the 2:1 MMR 
by using the 
FMA (Frequency Map Analysis) as described in Marzari et al.(2003).
The semimajor axes of Jupiter and Saturn correspond to values 
of the NICE model. Migration is switched off, so that the 
planets do not leave the resonance (frozen model). 
As pointed out above, the two resonance variables
$\theta_1$ and $\theta_2$
may both librate (apsidal corotation), or only one may librate while the other
circulates \citep{masc21}. 
We applied the FMA analysis for both cases.
Our results show that
the stability of Trojans depends strongly 
on their initial conditions and on the behaviour of the two resonance
arguments.
The upper diagram in Fig.~\ref{fig13a} represents
a diffusion portrait for Trojan orbits with Jupiter and Saturn in apsidal
corotation. Corotation is possible around $0^{\circ}$ or $180^{\circ}$.
Since we obtain in most migration models corotation around $180^{\circ}$,
we use this alignment mode for producing the diagram.
Extended
stability regions appear between medium and high inclinations and for 
a large range of
values for libration amplitudes $D$ of Trojans. 
Empty regions in the plot indicate instability times shorter than 
1 Myr. 
The most stable region (the red one)
has  values for diffusion speed 
comparable to those of present Jupiter Trojans
\citep{marscho} suggesting that bodies can survive for a long interval 
of time of the order of some Gyrs. 
For bodies with higher diffusion speed we still expect lifetimes 
of the order $10^7-10^8$ yrs. The stable region extends down 
to low inclinations with libration amplitudes of about $D \sim 
60^{\circ}$ where we found stability with different integrators, 
contrary to Michtchenko et al.(2001).

In the lower diagram of Fig.~\ref{fig13a} we consider a different dynamical state for
the two planets in resonance. Only one of the two critical arguments
librates. Consequently, $\Delta\varpi$ circulates. The stability area is 
more extended in this case and orbits with low inclination can be found 
at low values of libration amplitude $D$. These results 
reinforce the idea that corotation contributes significantly
to reduce dynamical lifetimes of Trojans.

\section{Conclusions}

We investigate the depletion of an alleged initial Jupiter Trojan
population in the frame of the NICE model 
describing the early migration
phase of the outer planets during which 
Jupiter and Saturn may have crossed their mutual 2:1 MMR. The loss of
an initial population, possibly trapped during the growth of the planet, 
is due to the synergy of three different effects:
\begin{itemize}
\item A secular resonance with the frequency 
$g_1$, one of the two frequencies that, according
to the Lagrange--Laplace theory, determine the 
the secular evolution of the eccentricity and perihelion
longitude of Jupiter and Saturn.
\item Secondary resonances due to commensurabilities between a critical
resonance argument of the 2:1 MMR and the libration frequency of the
critical argument of the Trojan orbits.
\item Jupiter and Saturn's apsidal corotation after the 2:1 MMR crossing 
\end{itemize}

While the planets approach the 2:1 MMR, the secular resonance $g_1$ 
sweeps through the Trojan region. It appears first
at high inclinations and it moves down to almost zero degrees when 
the planets reach the center
of the 2:1 MMR. It moves up again at higher inclinations after the resonance
crossing, sweeping for a second time the Trojan region. Also secondary
resonances appear before and after the 2:1 MMR crossing but they 
sweep across the Trojan region at a faster rate, in particular
after the 2:1 MMR crossing. 
Before the 2:1 MMR crossing, secondary resonances remove very few
Trojans while they are more effective after the crossing 
because of the increase of Jupiter's eccentricity.
Also the secular resonance $g_1$ is stronger after the 2:1 MMR crossing 
for the higher eccentricity of Jupiter 
and also because of the apsidal corotation of Jupiter 
and Saturn's orbit. 
When the frequency $g_1$ moves out of 
the Trojan region but is still bordering it, the 
secular term $g - g_1$ is
strong enough to 
perturb the Tojan motion causing instability
on a relatively longer timescale. 
While Trojans are removed, new Trojans can be captured 
by the reverse chaotic path from
the surrounding planetesimal population which drives planetary
migration. 
The newly captured Trojans might be lost again until the secular resonance, 
secondary resonances and higher order mean motion resonances between Jupiter and Saturn 
disappear. 

The center of the 2:1 MMR,
where at least one of the critical resonance arguments librates, 
is not particularly
effective in destabilizing Jupiter Trojans. 
Its effect is much weaker as compared
to the secular resonance $g_1$ and the secondary resonances after the
2:1 MMR crossing. When the planets are steadily locked in resonance 
we find extended 
stability regions in the phase space of Trojan orbits. 

\section*{Acknowledgments}

\end{document}